\documentclass[a4paper,10pt]{article}
\usepackage{graphicx}
\usepackage{comment}
\usepackage{amssymb}
\usepackage{amsmath}
\usepackage{nicefrac}
\usepackage{graphicx}
\usepackage{dcolumn}
\usepackage{bm}
\usepackage{comment}
\usepackage{multirow}
\usepackage{cite}

\begin{document}

\huge

\begin{center}
Expression of the Holtsmark function in terms of hypergeometric $_2F_2$ and Airy $\mathrm{Bi}$ functions
\end{center}

\vspace{0.5cm}

\large

\begin{center}
Jean-Christophe Pain\footnote{jean-christophe.pain@cea.fr}
\end{center}

\normalsize

\begin{center}
\it CEA, DAM, DIF, F-91297 Arpajon, France
\end{center}

\vspace{0.5cm}

\begin{abstract}
The Holtsmark distribution has applications in plasma physics, for the electric-microfield distribution involved in spectral line shapes for instance, as well as in astrophysics for the distribution of gravitating bodies. It is one of the few examples of a stable distribution for which a closed-form expression of the probability density function is known. However, the latter is not expressible in terms of elementary functions. In the present work, we mention that the Holtsmark probability density function can be expressed in terms of hypergeometric function $_2F_2$ and of Airy function of the second kind $\mathrm{Bi}$ and its derivative. The new formula is simpler than the one proposed by Lee involving $_2F_3$ and $_3F_4$ hypergeometric functions.
\end{abstract}

\section{Introduction}
\label{intro}

\noindent In 1919, Norwegian physicist J. Holtsmark proposed a distribution modelling the fluctuating fields in plasma due to chaotic motion of charged particles. The Holtsmark distribution of electric micro-fields in the quasi-static approximation is \cite{holtsmark19,hey76,hey89,salzmann,demura10,gigosos14}:

\begin{equation}
H(\beta)=\frac{2\beta}{\pi}\int_0^{\infty}x\sin(\beta x)\exp\left(-x^{3/2}\right)dx,
\end{equation}

\noindent which can also be written

\begin{equation}
H(\beta)=\frac{2}{\pi\beta}\int_0^{\infty}x\sin(x)\exp\left[-\left(\frac{x}{\beta}\right)^{3/2}\right]dx.
\end{equation}

\noindent The probability distribution of the ionic electric field in a plasma follows Holtsmark's distribution when the ionic correlations are weak. Such a distribution was also used for the distribution of celestial bodies \cite{chandra42,chandra43,pietronero02}. Holtsmark's distribution enables one to estimate the probability of realization of an atomic bound state in a plasma \cite{dyachkov98} and is encountered in the study of L\'evy processes \cite{barndorff-nielsen01}. In a paper about the statistics of the gravitational force in various dimensions of space, Chavanis recalls the main lines of the calculation of the distribution of the gravitational force in three dimensions, discuss the main mathematical properties of the Holtsmark distribution, and gives interesting references on the subject \cite{chavanis09}. In all the fields of physical modelling involving the Holtsmark distribution, the following function 

\begin{equation}\label{ints}
S(\beta)=\frac{1}{\pi}\int_0^{\infty}\cos(\beta x)\exp\left(-x^{3/2}\right)dx.
\end{equation}

\noindent plays an important role \cite{stam08,stam12,stam13}; it is related to the function $H(\beta)$ by

\begin{equation}
\frac{\partial S(\beta)}{\partial\beta}=-\frac{1}{2}\frac{H(\beta)}{\beta}.
\end{equation}

\noindent The function $S(\beta)$ is often called the Holtsmark function in mathematical statistics, but must not be confused with the Holtsmark distribution $H(\beta)$ used in physics and astronomy. $S(\beta)$ is in fact the probability density function of the Holtsmark distribution. It is often believed that it is not possible to obtain a rather compact expression of  $S(\beta)$ (or $H(\beta)$) in terms of known functions (see for instance page 183 of the book ``Atomic Physics in Hot Plasmas'', by D. Salzmann: ``There is no analytical solution for the integral in terms of known elementary or special functions'' \cite{salzmann}). There are many interesting works about the numerical computation of the Holtsmark distribution, for instance using Monte Carlo methods \cite{kozlitin11} or rational-fraction approximations \cite{poquerusse00}. The peak value of $S(\beta)$ is 

\begin{equation}
S(0)=\frac{2}{3\pi}\mathrm{\Gamma}\left(\frac{2}{3}\right)\approx 0.2874
\end{equation}

\noindent and the two following expansions are often used \cite{mihalas78,hummer86}:

\begin{eqnarray}\label{smallarg}
S(\beta)&=&\frac{2}{3\pi}\sum_{n=0}^{\infty}(-1)^n\mathrm{\Gamma}\left(\frac{4n+2}{3}\right)\frac{\beta^{2n}}{(2n)!},
\end{eqnarray}

\noindent where $\mathrm{\Gamma}$ is the usual Gamma function and

\begin{eqnarray}\label{bigarg}
S(\beta)&\approx&\frac{1}{\pi}\sum_{n=1}^{\infty}\frac{(-1)^{n+1}}{n!}\mathrm{\Gamma}\left(\frac{3n+2}{2}\right)\sin\left(\frac{3\pi n}{4}\right)\beta^{-(3n+2)/2}.
\end{eqnarray}

\noindent Equation (\ref{smallarg}) results from expansion series of $\cos(\beta x)$ and term-by-term integration. It is an absolutely convergent series;  and Eq. (\ref{bigarg}) results from expansion series of $\exp\left(-x^ {3/2}\right)$ and term-by-term integration. It is not an absolutely convergent series, but an asymptotic expansion \cite{dingle73} for $\beta >0$. It is worth mentioning that expansions (\ref{smallarg}) and (\ref{bigarg}) can be derived by applying successively the Barnes-Mellin transform \cite{hardy1916,prudnikov1,flajolet95,paris01,barnes05}:

\begin{equation}
\mathcal{M}(u)=\int_0^{\infty}\beta^{u-1}S(\beta)d\beta 
\end{equation}

\noindent and its inverse

\begin{equation}
S(\beta)=\frac{1}{2\pi i}\int_{c-i\infty}^{c+i\infty}\beta^ {-u}\mathcal{M}(u)du,
\end{equation}

\noindent where $0<c<1$, and calculating the latter integral in the complex plane. Equation (\ref{smallarg}) is useful in order to approximate $S(\beta)$ for small values of $\beta$:

\begin{equation}
S(\beta)\approx 0.2874-0.1061~\beta^2+0.0246~\beta^4+\cdots\;\;\;\;\mathrm{for}\;\;\;\;\beta\ll 1,
\end{equation}

\noindent where $\mathrm{\Gamma}$ is the usual Gamma function and Eq. (\ref{bigarg}) provides an approximant of $S(\beta)$ for large vales of $\beta$:

\begin{equation}
S(\beta)\asymp 0.2992~\beta^{-5/2}+0.9549~\beta^{-4}+1.9635~\beta^{-11/2}+\cdots\;\;\;\;\mathrm{for}\;\;\;\;\beta\gg 1.
\end{equation}

\begin{figure}[!ht]
\begin{center}
\vspace{5mm}
\includegraphics[width=8cm]{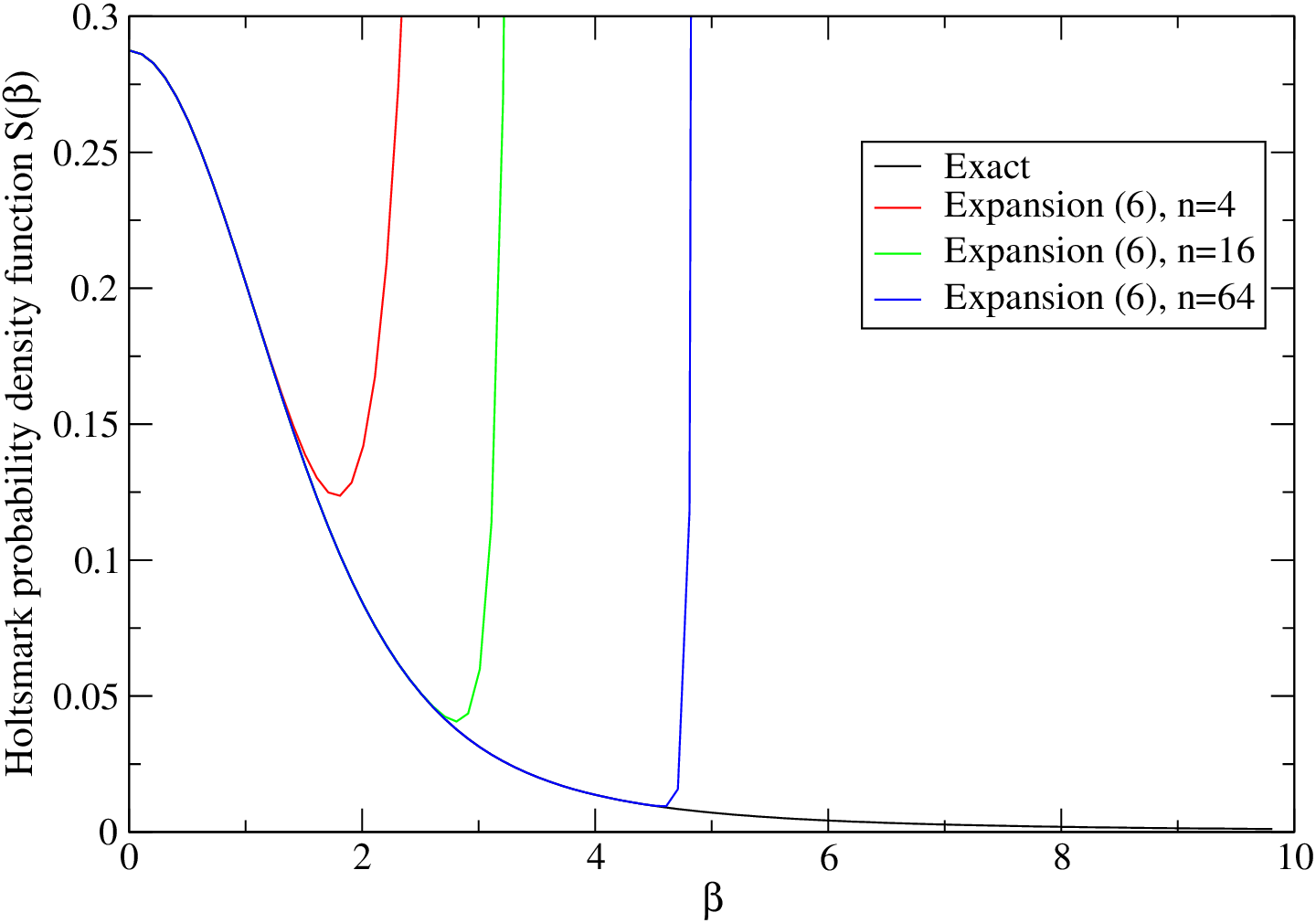}
\vspace{5mm}
\end{center}
\caption{Expansion (\ref{smallarg}) for different orders: 4 (red curve), 16 (green curve) and 64 (blue curve) compared to the exact probability density function $S(\beta)$ (black curve).}\label{fig_holts}
\end{figure}

\begin{figure}[!ht]
\begin{center}
\vspace{5mm}
\includegraphics[width=8cm]{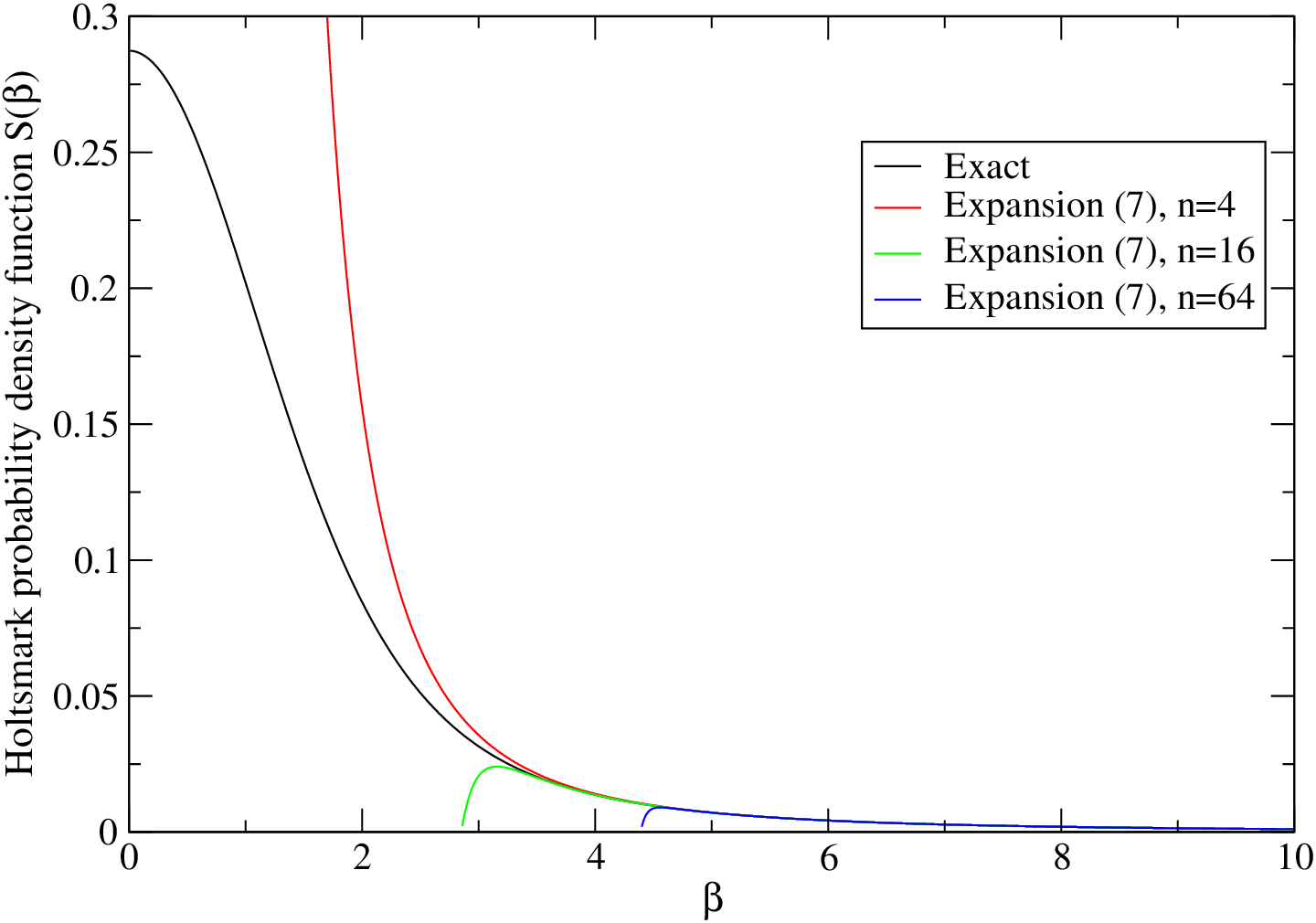}
\vspace{5mm}
\end{center}
\caption{Expansion (\ref{bigarg}) for different orders: 4 (red curve), 16 (green curve) and 64 (blue curve) compared to the exact probability density function $S(\beta)$ (black curve).}\label{fig_holtl}
\end{figure}

\noindent Expansions (\ref{smallarg}) and (\ref{bigarg}) are represented in Figs. (\ref{fig_holts}) and (\ref{fig_holtl}) respectively for different orders (4, 16 and 64) and compared to the exact probability density function $S(\beta)$. The Fourier transform of $S(\beta)$ is

\begin{eqnarray}\label{fou}
\tilde{S}(\tau)&=&\int_{-\infty}^{\infty}S(\beta)\exp\left(-i\beta\tau\right)d\beta\nonumber\\
&=&\frac{1}{2\pi}\int_{-\infty}^{\infty}\left[\int_0^{\infty}\exp\left(-x^{3/2}\right)\left\{\exp\left[i\beta(x-\tau)\right]+\exp\left[i\beta(x+\tau)\right]\right\}dx\right]d\beta\nonumber\\
&=&\exp(-|\tau|^{3/2}).
\end{eqnarray}

\noindent In the following section, we give an expression of the Holtsmark probability density function $S(\beta)$ and distribution $H(\beta)$ in terms of special functions, and in particular in terms of hypergeometric $_2F_2$ and Airy $\mathrm{Bi}$ functions. 

\section{Analytical expressions of $S(\beta)$ and $H(\beta)$}

\noindent It was found by Lee \cite{lee10} that $S(\beta)$ can be expressed in terms of hypergeometric functions $_2F_3$ and $_3F_4$:  

\begin{eqnarray}\label{nc}
S(\beta)&=&\frac{\mathrm{\Gamma}(5/3)}{\pi}~_2F_3
\left(
\begin{array}{c}
\frac{5}{12}, \frac{11}{12}\\
\frac{1}{3}, \frac{1}{2}, \frac{5}{6}\\
\end{array};-\frac{4\beta^6}{3^6}
\right)\nonumber\\
& &-\frac{\beta^2}{3\pi}~_3F_4
\left(
\begin{array}{c}
\frac{3}{4}, 1, \frac{5}{4}\\
\frac{2}{3}, \frac{5}{6}, \frac{7}{6}, \frac{4}{3}\\
\end{array}; -\frac{4\beta^6}{3^6}
\right)
\nonumber\\
& &+\frac{7\beta^4\mathrm{\Gamma}(4/3)}{3^4\pi}~_2F_3
\left(
\begin{array}{c}
\frac{13}{12}, \frac{19}{12}\\
\frac{7}{6}, \frac{3}{2}, \frac{5}{3}\\
\end{array}; -\frac{4\beta^6}{3^6}
\right).
\end{eqnarray}

\noindent Such an expression was also mentioned by $\mathrm{\c{C}}$opuro$\mathrm{\check{g}}$lu \cite{copuroglu19} and Mehmeto$\mathrm{\check{g}}$lu. We provide an expression which is simpler than Eq. (\ref{nc}), in the sense that it involves hypergeometric functions with lower indices (lower numbers of parameters) and Airy (or Bessel) functions. We give below the main steps of two different ways to obtain the new formula. The first one consists in using inverse Fourier transform of $\exp(-|\tau|^{3/2})$ \cite{stam08,stam12,stam13}:

\begin{equation}\label{inv}
S(\beta)=\frac{1}{2\pi}\int_{-\infty}^{\infty}\exp\left(-|\xi|^{3/2}-i\beta\xi\right)d\xi
\end{equation}

\noindent and splitting the integral in two parts (one from $-\infty$ to 0 and the other one from 0 to $\infty$). For instance, considering

\begin{equation}
I(\beta)=\int_0^{\infty}e^{-t^{3/2}}e^{-i\beta t}dt,
\end{equation}

\noindent we find, with the change of variable $v=t^{3/2}$:

\begin{equation}
I(\beta)=\frac{2}{3}\int_0^{\infty}e^{-v}v^{-1/3}e^{-i\beta v^{2/3}}dv,
\end{equation}

\noindent which yields, after expanding $e^{-i\beta v^{2/3}}$ in series:

\begin{equation}
I(\beta)=\frac{2}{3}\sum_{k=0}^{\infty}\frac{(-i\beta)^k}{k!}\int_0^{\infty}v^{\frac{(2k-1)}{3}}e^{-v}dv=\frac{2}{3}\sum_{k=0}^{\infty}\frac{(-i\beta)^k}{k!}\mathrm{\Gamma}\left(\frac{2(k+1)}{3}\right).
\end{equation}

\noindent Then, knowing that the general form of a generalized hypergeometric series $_pF_q$ is

\begin{eqnarray}\label{def_fpq}
_pF_q\left[\begin{array}{c}
\alpha_1\cdots\alpha_p\\
\gamma_1,\cdots,\gamma_q
\end{array};z\right]&=&\sum_{n=0}^{\infty}\frac{\prod_{i=1}^p(\alpha_i)_n}{\prod_{j=1}^q(\gamma_j)_n}\frac{z^n}{n!},
\end{eqnarray}

\noindent where $(\lambda)_n$ represents the Pochhammer symbol

\begin{equation}
(\lambda)_n=\lambda(\lambda+1)\cdots(\lambda+n-1)=\frac{\mathrm{\Gamma}(\lambda+n)}{\mathrm{\Gamma}(\lambda)},
\end{equation}

\noindent one gets, using a computer algebra system \cite{mathematica}:

\begin{eqnarray}
I(\beta)&=&\frac{2}{3}\mathrm{\Gamma}\left(\frac{2}{3}\right)~_1F_1\left(\begin{array}{c}
\frac{5}{6}\\
\frac{2}{3}
\end{array}; \frac{4i\beta^3}{27}\right)-\frac{2i\beta}{9}\mathrm{\Gamma}\left(\frac{1}{3}\right)~_1F_1\left(\begin{array}{c}
\frac{7}{6}\\
\frac{4}{3}
\end{array}; \frac{4i\beta^3}{27}\right)\nonumber\\
& &-\frac{\beta^2}{3}~_2F_2\left(\begin{array}{c}
1,\frac{3}{2}\\
\frac{4}{3}, \frac{5}{3}
\end{array}; \frac{4i\beta^3}{27}\right).
\end{eqnarray}

\noindent In the same way, we have

\begin{eqnarray}
J(\beta)&=&\int_{-\infty}^{0}e^{-|t|^{3/2}}e^{-i\beta t}dt\nonumber\\
&=&\frac{2}{3}\mathrm{\Gamma}\left(\frac{2}{3}\right)~_1F_1\left(\begin{array}{c}
\frac{5}{6}\\
\frac{2}{3}
\end{array};-\frac{4i\beta^3}{27}\right)+\frac{2i\beta}{9}\mathrm{\Gamma}\left(\frac{1}{3}\right)~_1F_1\left(\begin{array}{c}
\frac{7}{6}\\
\frac{4}{3}
\end{array};-\frac{4i\beta^3}{27}\right)\nonumber\\
& &-\frac{\beta^2}{3}~_2F_2\left(\begin{array}{c}
1,\frac{3}{2}\\
\frac{4}{3}, \frac{5}{3}
\end{array};-\frac{4i\beta^3}{27}\right).
\end{eqnarray}

\noindent In the present case, $\beta$ is real, the argument of the $_1F_1$ and $_2F_2$ functions is pure imaginary and the sum of two $_1F_1$ (or two $_2F_2$) functions with opposite imaginary arguments is real. Therefore, $I(\beta)$ and $J(\beta)$ are complex conjugates, and their sum is real. Since Airy function $\mathrm{Bi}$ and its derivative $\mathrm{Bi}'$ \cite{abramowitz,ghatak91} can be expressed using confluent hypergeometric functions $_1F_1$ \cite{dmlf}:

\begin{eqnarray}
\mathrm{Bi}(x)&=&\frac{1}{3^{1/6}\mathrm{\Gamma}\left(\frac{2}{3}\right)}\exp\left(-\frac{2}{3}x^{3/2}\right)~_1F_1\left(\begin{array}{c}
\frac{1}{6}\\
\frac{1}{3}
\end{array};\frac{4}{3}x^{3/2}\right)\nonumber\\
& &+\frac{3^{5/6}}{2^{2/3}\mathrm{\Gamma}\left(\frac{1}{3}\right)}\left(\frac{2}{3}\right)^{2/3}x\exp\left(-\frac{2}{3}x^{3/2}\right)~_1F_1\left(\begin{array}{c}
\frac{5}{6}\\
\frac{5}{3}
\end{array};\frac{4}{3}x^{3/2}\right)
\end{eqnarray}

\noindent and

\begin{eqnarray}
\mathrm{Bi}'(x)&=&\frac{3^{1/6}}{\mathrm{\Gamma}\left(\frac{1}{3}\right)}\exp\left(-\frac{2}{3}x^{3/2}\right)~_1F_1\left(\begin{array}{c}
-\frac{1}{6}\\
-\frac{1}{3}
\end{array};\frac{4}{3}x^{3/2}\right)\nonumber\\
& &+\frac{1}{2\times 3^{1/6}\mathrm{\Gamma}\left(\frac{2}{3}\right)}x^2\exp\left(-\frac{2}{3}x^{3/2}\right)~_1F_1\left(\begin{array}{c}
\frac{7}{6}\\
\frac{7}{3}
\end{array};\frac{4}{3}x^{3/2}\right),
\end{eqnarray}

\noindent it is possible, with the help of a computer algebra system \cite{mathematica}, to recognize

\begin{eqnarray}\label{final}
S(\beta)&=&\frac{-\beta^2}{6\pi}\left[~_2F_2\left(\begin{array}{c}
1,\frac{3}{2}\\
\frac{4}{3}, \frac{5}{3}
\end{array}; -\frac{4i\beta^3}{27}\right)+~_2F_2\left(\begin{array}{c}
1,\frac{3}{2}\\
\frac{4}{3}, \frac{5}{3}
\end{array}; \frac{4i\beta^3}{27}\right)\right]\nonumber\\
& &+\frac{4}{3\times 3^{2/3}}\left[\mathrm{Bi}'\left(-\frac{\beta^2}{3\times 3^{1/3}}\right)\cos\left(\frac{2\beta^3}{27}\right)\right.\nonumber\\
& &\left.+\frac{\beta}{3^{2/3}}~\mathrm{Bi}\left(-\frac{\beta^2}{3\times 3^{1/3}}\right)\sin\left(\frac{2\beta^3}{27}\right)\right].
\end{eqnarray}

\noindent The arguments of the $_2F_2$ functions are pure imaginary complex numbers, but the sum of the two functions is real. For $x$ positive, the function $\mathrm{Bi}(-x)$ is related to the Bessel functions of fractional order $J_{-1/3}$ and $J_{1/3}$ and its derivative to the Bessel functions of fractional order $J_{-2/3}$ and $J_{2/3}$ \cite{abramowitz,prudnikov}. One has therefore 

\begin{eqnarray}
S(\beta)&=&\frac{4\beta^2}{27\sqrt{3}}\left\{\cos\left(\frac{2\beta^3}{27}\right)\left[J_{-2/3}\left(\frac{2\beta^3}{27}\right)+J_{2/3}\left(\frac{2\beta^3}{27}\right)\right]\right.\nonumber\\
& &\left.+\sin\left(\frac{2\beta^3}{27}\right)\left[J_{-1/3}\left(\frac{2\beta^3}{27}\right)-J_{1/3}\left(\frac{2\beta^3}{27}\right)\right]\right\}\nonumber\\
& &-\frac{\beta^2}{6\pi}\left[~_2F_2\left(\begin{array}{c}
1,\frac{3}{2}\\
\frac{4}{3}, \frac{5}{3}
\end{array}; -\frac{4i\beta^3}{27}\right)+~_2F_2\left(\begin{array}{c}
1,\frac{3}{2}\\
\frac{4}{3}, \frac{5}{3}
\end{array}; \frac{4i\beta^3}{27}\right)\right].
\end{eqnarray}

\noindent Another way to derive Eq. (\ref{final}) is to start from Eq. (\ref{smallarg}), which reads equivalently

\begin{equation}
S(\beta)=\frac{2}{3\pi\beta}\sum_{m=0}^{\infty}(-1)^m\beta^{2m+1}\frac{\mathrm{\Gamma}\left(\frac{2(2m+1)}{3}\right)}{\mathrm{\Gamma}(2m+1)},
\end{equation}

\noindent and setting $m=3p+q$, we get

\begin{equation}
S(\beta)=\frac{2}{3\pi\beta}\sum_{p=0}^{\infty}\sum_{q=0}^2(-1)^{3p+q}\beta^{6p+2q+1}\frac{\mathrm{\Gamma}\left(4p+\frac{2(2q+1)}{3}\right)}{\mathrm{\Gamma}(6p+2q+1)},
\end{equation}

\noindent which becomes, with $r=2p$:

\begin{equation}
S(\beta)=\frac{2}{3\pi\beta}\sum_{r=0}^{\infty}\sum_{q=0}^2\left[\frac{1+(-1)^r}{2}\right]e^{i\pi\frac{3r}{2}}(-1)^q\beta^{3r+2q+1}\frac{\mathrm{\Gamma}\left(2r+\frac{2(2q+1)}{3}\right)}{\mathrm{\Gamma}(3r+2q+1)}.
\end{equation}

\noindent Using Gauss' multiplication formula, it is possible to write

\begin{equation}
\frac{\mathrm{\Gamma}\left(2r+\frac{2(2q+1)}{3}\right)}{\mathrm{\Gamma}(3r+2q+1)}=\sqrt{2\pi}\frac{2^{2r+\frac{4q}{3}+\frac{1}{6}}}{3^{3r+2q+\frac{1}{2}}}\frac{\mathrm{\Gamma}\left(\frac{2q}{3}+\frac{5}{6}\right)\left(\frac{2q}{3}+\frac{5}{6}\right)_r}{\prod_{k=1}^2\mathrm{\Gamma}\left(\frac{2q+1+k}{3}\right)\left(\frac{2q+1+k}{3}\right)_r},
\end{equation}

\noindent which leads to

\begin{eqnarray}
S(\beta)&=&\frac{2^{7/6}}{3\sqrt{6\pi}}\sum_{q=0}^2(-1)^q\left(\frac{4\beta^3}{27}\right)^{\frac{2q}{3}}\frac{\mathrm{\Gamma}\left(\frac{4q+5}{6}\right)}{\prod_{k=1}^2\mathrm{\Gamma}\left(\frac{2q+1+k}{3}\right)}\nonumber\\
& &\times\left\{~_2F_2\left(\begin{array}{c}
1,\frac{4q+5}{6}\\
\frac{2q+2}{3}, \frac{2q+3}{3}
\end{array}; \frac{4i\beta^3}{27}\right)+~_2F_2\left(\begin{array}{c}
1,\frac{4q+5}{6}\\
\frac{2q+2}{3}, \frac{2q+3}{3}
\end{array}; -\frac{4i\beta^3}{27}\right)\right\}.\nonumber\\
& &
\end{eqnarray}

\noindent The latter expression can be simplified into

\begin{eqnarray}
S(\beta)&=&\frac{2^{7/6}}{3\sqrt{6\pi}}\left\{\frac{\mathrm{\Gamma}\left(\frac{5}{6}\right)}{\mathrm{\Gamma}\left(\frac{2}{3}\right)}\left[~_2F_2\left(\begin{array}{c}
1,\frac{5}{6}\\
\frac{2}{3}, 1
\end{array};-\frac{4i\beta^3}{27}\right)+~_2F_2\left(\begin{array}{c}
1,\frac{5}{6}\\
\frac{2}{3}, 1
\end{array};\frac{4i\beta^3}{27}\right)\right]\right.\nonumber\\
& &-\left(\frac{4\beta^3}{27}\right)^{2/3}\frac{\mathrm{\Gamma}\left(\frac{3}{2}\right)}{\mathrm{\Gamma}\left(\frac{4}{3}\right)\mathrm{\Gamma}\left(\frac{5}{3}\right)}\left[~_2F_2\left(\begin{array}{c}
1,\frac{3}{2}\\
\frac{4}{3}, \frac{5}{3}
\end{array};-\frac{4i\beta^3}{27}\right)+~_2F_2\left(\begin{array}{c}
1,\frac{3}{2}\\
\frac{4}{3}, \frac{5}{3}
\end{array};\frac{4i\beta^3}{27}\right)\right]\nonumber\\
& &+\left.\left(\frac{4\beta^3}{27}\right)^{2/3}\frac{\mathrm{\Gamma}\left(\frac{13}{6}\right)}{\mathrm{\Gamma}\left(\frac{7}{3}\right)}\left[~_2F_2\left(\begin{array}{c}
1,\frac{13}{6}\\
2, \frac{7}{3}
\end{array};-\frac{4i\beta^3}{27}\right)+~_2F_2\left(\begin{array}{c}
1,\frac{13}{6}\\
2, \frac{7}{3}
\end{array};\frac{4i\beta^3}{27}\right)\right]\right\}\nonumber\\
& &
\end{eqnarray}

\noindent and since

\begin{equation}
~_2F_2\left(\begin{array}{c}
1, \frac{5}{6}\\
\frac{2}{3}, 1
\end{array};x\right)=~_1F_1\left(\begin{array}{c}
\frac{5}{6}\\
\frac{2}{3}
\end{array};x\right)
\end{equation}

\noindent as well as

\begin{equation}
~_2F_2\left(\begin{array}{c}
1,\frac{13}{6}\\
2, \frac{7}{3}
\end{array};x\right)=\frac{8}{7x}\left[~_1F_1\left(\begin{array}{c}
\frac{7}{6}\\
\frac{4}{3}
\end{array};x\right)-1\right],
\end{equation}

\noindent we obtain, with the help of a computer algeba system \cite{mathematica}, Eq. (\ref{final}). Using the relations 

\begin{equation}
\frac{\partial}{\partial x}~_2F_2(a_1,a_2;b_1,b_2;x)=\frac{a_1a_2}{b_1b_2}~_2F_2(a_1+1,a_2+1;b_1+1,b_2+1;x)
\end{equation}

\noindent and $\mathrm{Bi}''(x)=x~\mathrm{Bi}(x)$, we find for the Holtsmark distribution

\begin{eqnarray}
H(\beta)&=&\frac{2\beta^2}{3\pi}\left[~_2F_2\left(\begin{array}{c}
1,\frac{3}{2}\\
\frac{4}{3}, \frac{5}{3}
\end{array}; -\frac{4i\beta^3}{27}\right)+~_2F_2\left(\begin{array}{c}
1,\frac{3}{2}\\
\frac{4}{3}, \frac{5}{3}
\end{array}; \frac{4i\beta^3}{27}\right)\right]\nonumber\\
& &-\frac{i\beta^5}{10\pi}\left[~_2F_2\left(\begin{array}{c}
2,\frac{5}{2}\\
\frac{7}{3}, \frac{8}{3}
\end{array}; -\frac{4i\beta^3}{27}\right)-~_2F_2\left(\begin{array}{c}
2,\frac{5}{2}\\
\frac{7}{3}, \frac{8}{3}
\end{array}; \frac{4i\beta^3}{27}\right)\right]\nonumber\\
& &-\frac{8\beta}{3^4\times 3^{2/3}}\left\{3^{1/3}\mathrm{Bi}\left(-\frac{\beta^2}{3\times 3^{1/3}}\right)\left[4\beta^3\cos\left(\frac{2\beta^3}{27}\right)+9\sin\left(\frac{2\beta^3}{27}\right)\right]\right.\nonumber\\
& &\left.-12\beta^2\mathrm{Bi'}\left(-\frac{\beta^2}{3\times 3^{1/3}}\right)\sin\left(\frac{2\beta^3}{27}\right)\right\},
\end{eqnarray}

\noindent or, in terms of Bessel functions:

\begin{eqnarray}
H(\beta)&=&\frac{2\beta^2}{3\pi}\left[~_2F_2\left(\begin{array}{c}
1,\frac{3}{2}\\
\frac{4}{3}, \frac{5}{3}
\end{array}; -\frac{4i\beta^3}{27}\right)+~_2F_2\left(\begin{array}{c}
1,\frac{3}{2}\\
\frac{4}{3}, \frac{5}{3}
\end{array}; \frac{4i\beta^3}{27}\right)\right]\nonumber\\
& &-\frac{i\beta^5}{10\pi}\left[~_2F_2\left(\begin{array}{c}
2,\frac{5}{2}\\
\frac{7}{3}, \frac{8}{3}
\end{array}; -\frac{4i\beta^3}{27}\right)-~_2F_2\left(\begin{array}{c}
2,\frac{5}{2}\\
\frac{7}{3}, \frac{8}{3}
\end{array}; \frac{4i\beta^3}{27}\right)\right]\nonumber\\
& &-\frac{8\beta^2}{3^5\times \sqrt{3}}\left\{\left[J_{-1/3}\left(\frac{2\beta^3}{27}\right)-J_{1/3}\left(\frac{2\beta^3}{27}\right)\right]\right.\nonumber\\
& &\times\left[4\beta^3\cos\left(\frac{2\beta^3}{27}\right)+9\sin\left(\frac{2\beta^3}{27}\right)\right]\nonumber\\
& &\left.-4\beta^3\left[J_{-2/3}\left(\frac{2\beta^3}{27}\right)+J_{2/3}\left(\frac{2\beta^3}{27}\right)\right]\sin\left(\frac{2\beta^3}{27}\right)\right\}.
\end{eqnarray}

\section{Conclusion}

We have provided an expression of the Holtsmark probability density function in terms of hypergeometric $_2F_2$ and Airy $\mathrm{Bi}$ and $\mathrm{Bi}'$ functions. The part involving $\mathrm{Bi}$ and $\mathrm{Bi}'$ functions can also be expressed in terms of Bessel functions of fractional order $J_{-2/3}$, $J_{-1/3}$, $J_{1/3}$ and $J_{2/3}$. The interest of the new expression (\ref{final}) for numerical evaluations is not obvious. It seems much more straightforward to base evaluation of the function on its standard series (\ref{smallarg}) and asymptotic expansion (\ref{bigarg}), which have a very simple form. The difficulty of expressing the Holtsmark distribution in terms of standard special functions is only due to the argument of the gamma function being a fraction of the summation index in the expansion (\ref{smallarg}) and (\ref{bigarg}). The new formulation may yield the discovery of further relations, in particular using identities for Airy and $_2F_2$ functions. It may also help to study asymptotic behaviours and derive approximants of the Holtsmark distribution.


\begin{thebibliography}{}

\bibitem{holtsmark19} J. Holtsmark, {\it Uber die Verbreiterung von Spektrallinien}, Ann. Phys. (Leipzig) {\bf 58}, 577-630 (1919).

\bibitem{hey76} J. D. Hey, {\it A generalization of some functions of Holtsmark, and Chandrasekhar and Von Neumann}, J. Quant. Spectrosc. Radiat. Transfer {\bf 16}, 947-931 (1976). 

\bibitem{hey89} J. D. Hey, {\it Further properties of the generalized functions of Holtsmark, and Chandrasekhar and Von Neumann}, J. Quant. Spectrosc. Radiat. Transfer {\bf 41}, 167-171 (1989).

\bibitem{salzmann} D. Salzmann, {\it Atomic Physics in Hot Plasmas}, International Series of Monographs on Physics, (Oxford University Press, 1998). 

\bibitem{demura10} A. V. Demura, {\it Physical Models of Plasma Microfield}, Int. J. of Spectrosc. {\bf 671073}, 42 pages (2010). 

\bibitem{gigosos14} M. A. Gigosos, {\it Stark broadening models for plasma diagnostics}, J. Phys. D: Appl. Phys. {\bf 47}, 343001 (2014).
 
\bibitem{chandra42} S. Chandrasekhar, {\it The Statistics of the Gravitational Field Arising from a Random Distribution of Stars. I. The Speed of Fluctuations}, Astrophys. J. {\bf 95}, 489-531 (1942).

\bibitem{chandra43} S. Chandrasekhar, {\it Stochastic Problems in Physics and Astronomy}, Rev. Mod. Phys. {\bf 15}, 1-89 (1943).

\bibitem{pietronero02} L. Pietronero, M. Bottaccio, R. Mohayaee and M. Montuori, {\it The Holtsmark distribution of forces and its role in gravitational clustering}, J. Phys.: Condens. Matter {\bf 14}, 2141\u20ac\u201c2152 (2002).

\bibitem{dyachkov98} L. G. D'yachkov, {\it Approximation for the probabilities of the realization of atomic bound states in a plasma}, J. Quant. Spectrosc. Radiat. Transfer {\bf 59}, 65-69 (1998).

\bibitem{barndorff-nielsen01} O. E. Barndorff-Nielsen, T. Mikosch and S. I. Resnick, {\it L\'evy Processes: Theory and Applications} (Springer, 2001).

\bibitem{chavanis09} P. H. Chavanis, {\it Statistics of the gravitational force in various dimensions of space: from Gaussian to L\'evy laws}, Eur. Phys. J. B {\bf 70}, 413-433 (2009).

\bibitem{stam08} E. Stambulchik, {\it Stark effect of high-$n$ hydrogen-like transitions: quasi-contiguous approximation}, J. Phys. B: At. Mol. Opt. Phys. {\bf 41}, 095703 (2008). 

\bibitem{stam12} E. Stambulchik and Y. Maron, {\it Quasi-Contiguous Approximation for Line-Shape Modeling in Plasmas}, AIP Conf. Proc. {\bf 1438}, 203-209 (2012).

\bibitem{stam13} E. Stambulchik and Y. Maron, {\it Quasicontiguous frequency-fluctuation model for calculation of hydrogen and hydrogenlike Stark-broadened line shapes in plasmas}, Phys. Rev. E {\bf 87}, 053108 (2013).

\bibitem{kozlitin11} I. A. Kozlitin, {\it Simulating the Holtsmark distribution by the Monte Carlo method}, Mathematical Models and Computer Simulations {\bf 3}, 58-64 (2011).

\bibitem{poquerusse00} A. Poqu\'erusse, {\it Simplified rational approximations for Holtsmark and related functions}, Eur. Phys. J. D {\bf 10}, 307-308 (2000).

% developpements asymptotiques
\bibitem{mihalas78} D. Mihalas, {\it Stellar Atmospheres}, 2nd edn. (San Francisco: Freeman, 1978).

\bibitem{hummer86} D. G. Hummer, {\it Rational approximations for the Holtsmark distribution, its cumulative and derivative}, J. Quant. Spectrosc. Radiat. Transfer {\bf 36}, 1-5 (1986). 

\bibitem{dingle73} R. B. Dingle, {\it Asymptotic Expansions: Their Derivation and Interpretation} (Academic, London, 1973).

\bibitem{hardy1916} G. H. Hardy and J. E. Littlewood, {\it Contributions to the Theory of the Riemann Zeta-Function and the Theory of the Distribution of Primes}, Acta Mathematica, 41(1916) pp.119-196. (See notes therein for further references to Cahen's and Mellin's work, including Cahen's thesis.)

\bibitem{prudnikov1} A. P. Prudnikov, Yu. A. Brychkov and O. I. Marichev, {\it Evaluation of integrals and the Mellin transform}, Itogi Nauki i Tekhn. Ser. Mat. Anal. {\bf 27}, VINITI, Moscow,  3-146 (1989); J. Soviet Math. {\bf 54}, 1239-1341 (1991).

\bibitem{flajolet95} P. Flajolet, X. Gourdon and P. Dumas, {\it Mellin transforms and asymptotics: Harmonic sums}, Theoretical Computer Science. {\bf 144}, 3-58 (1995). 

\bibitem{paris01} R. B. Paris and D. Kaminsky, {\it Asymptotics and Mellin-Barnes Integrals} (Cambridge University Press, 2001).

\bibitem{barnes05} E. W. Barnes, {\it The Maclaurin sum formula}, Proc. London Math. Sot. {\bf 2}, 253-272 (1905).

\bibitem{lee10} W. H. Lee, {\it Continuous and Discrete Properties of Stochastic Processes}, PhD thesis (University of Nottingham, 2010), pp. 37-39. 

\bibitem{copuroglu19} E. $\mathrm{\c{C}}$opuro$\mathrm{\check{g}}$lu and T. Mehmeto$\mathrm{\check{g}}$lu, {\it Analytical evaluation of Holtsmark distribution of energies and its role in plasma microfields}, Journal of Science and Arts {\bf 2}, 523-528 (2019).

\bibitem{mathematica} Wolfram Research, Inc., Mathematica, Version 11.3, Champaign, IL (2018).

\bibitem{abramowitz} M. Abramowitz and I. A. Stegun, {\it Handbook of Mathematical Functions}, ninth printing, tenth GPO printing (Dover publications, Inc., New York, 1972).

\bibitem{ghatak91} A. K. Ghatak, R. L. Gallawa and I. C. Goyal, {\it  Modified Airy Function and WKB Solutions to the Wave Equation},
NIST Monograph 176 (U.S. Government Printing Office, Washington, 1991).

\bibitem{dmlf} NIST Digital Library of Mathematical Functions. http://dlmf.nist.gov/9.6, Release 1.0.24 of 2019-09-15. F. W. J. Olver, A. B. Olde Daalhuis, D. W. Lozier, B. I. Schneider, R. F. Boisvert, C. W. Clark, B. R. Miller, B. V. Saunders, H. S. Cohl, and M. A. McClain, eds., Eqs. 9.6.25 and 9.6.26. 

\bibitem{prudnikov} A. P. Prudnikov, Yu. A. Brychkov and O. I. Marichev, {\it Integrals and series}, Vol. 3: More Special Functions (Gordon and Breach, Amsterdam-Paris-New-York, 1986), translated from the Russian: {\it Integraly i ryady} by N. M. Queen.
 
\end{thebibliography}
\end{document}